\documentclass [12pt]{article}

\newcommand{\beq}{ \begin{eqnarray} }
\newcommand{\eeq}{ \end{eqnarray} }
\newcommand{\beqstar}{ \begin{eqnarray*} }
\newcommand{\eeqstar}{ \end{eqnarray*} }
\newcommand{\gsim}{ \mathop{}_{\textstyle \sim}^{\textstyle >} }
\newcommand{\lsim}{ \mathop{}_{\textstyle \sim}^{\textstyle <} }

\begin{document}
\baselineskip 0.7cm

\begin{titlepage}

\begin{center}

\hfill KEK-TH-725\\
\hfill CERN-TH/2000-338\\
\hfill YITP-00-62\\
\hfill hep-ph/0011216\\
\hfill \today

  {\large  Neutralino Warm Dark Matter}
  \vskip 0.5in {\large
    Junji~ Hisano$^{(a,b)}$,
    Kazunori~ Kohri$^{(c)}$ and
    Mihoko M.~Nojiri$^{(c)}$}
\vskip 0.4cm 
{\it 
(a) Theory Group, KEK, Oho 1-1, Tsukuba, Ibaraki 305-0801, Japan
}
\\
{\it 
(b) TH-devision, CERN, 1211 Geneva 23, Switzerland
}
\\
{\it 
(c) YITP, Kyoto University, Kyoto 606-8502, Japan
}
\vskip 0.5in

\abstract { In the supersymmetric (SUSY) standard model, the lightest
neutralino may be the lightest SUSY particle (LSP), and it is is a
candidate of the dark matter in the universe.  The LSP dark matter
might be produced by the non-thermal process such as heavy particle
decay after decoupling of the thermal relic LSP. If the produced LSP
is relativistic, and does not scatter enough in the thermal bath, the
neutralino LSP may contribute as the warm dark matter (WDM) to wash
out the small scale structure of $O(0.1)$ Mpc. In this letter we
calculate the energy reduction of the neutralino LSP in the thermal
bath and study whether the LSP can be the WDM. If temperature of the
production time $T_I$ is smaller than 5MeV, the bino-like LSP can be
the WDM and may contribute to the small-scale structure of $O(0.1)$
Mpc. The Higgsino-like LSP might also work as the WDM if $T_I<$
2MeV. The wino-like LSP cannot be the WDM in the favoured parameter
region.  }
\end{center}
\end{titlepage}
\setcounter{footnote}{0}
Existence of the dark matter in the universe is one of the important
observations for both cosmology and particle physics. The
supersymmetric standard model (SUSY SM) provides good candidates for
the dark matter in universe, since the R parity stabilizes the
lightest SUSY particle (LSP) \cite{DMreview}. In the supergravity
scenario the lightest neutralino with mass above O(50)GeV is the
LSP. It is produced in the thermal processes of the early universe,
and works as the cold dark matter (CDM) which explains the large-scale
structure in the universe well \cite{CDM}.

On the other hand, it is not yet clear whether the neutralino is
consistent with the small-scale structure formation of the
universe. It is pointed out that the CDM tends to make cuspy
structures in the halo density profiles \cite{cusp}. Such cuspy
structures may have drastic consequences to future observations in the
dark matter search \cite{moore}. On the other hand, it is argued that
the neutralino cuspy profile might be inconsistent with the radio
emission from centre of the galaxy \cite{gondolo}, although detailed
studies are waited for to confirm it. The consistency between the
observed structure of sub-galactic scale or cluster of galaxy
\cite{obs} and the numerical simulation of N body system \cite{nbody}
has been discussed extensively at present. The inflation models with
the small-scale perturbation suppressed \cite{inflation} and some new
candidates of the dark matter \cite{Kaplinghat:2000vt}\cite{Lin:2000qq}
have been proposed in order to explain it well.

Here we consider the neutralino dark matter produced by non-thermal
processes. If the neutralino dark matter is produced after decoupling
of the thermal relic neutralino, such neutralinos might remain without
annihilating and contribute as the dark matter.  The thermal relic of
the LSP on the other hand may be washed away by the entropy production
associated with the non-thermal production. Such a situation is
realized in some models, those are, the heavier moduli decay with mass
of the order of $10 \sim 100$ TeV at the late time
\cite{moduli}\cite{mr}, evaporation of cosmological defects
\cite{Lin:2000qq}, and so on. In those cases the produced LSP can be
highly relativistic compared with the thermal background.  If the LSP
keeps most of its energy from the scattering processes in the thermal
bath till the matter-radiation equality, the LSP behaves as the warm
dark matter (WDM). The small-scale structure in the universe within
the comoving free-streaming scale at the matter-radiation equality is
washed out, and the cuspy profiles in the halo would not be formed
\cite{Lin:2000qq}.

If the reduction of the LSP energy by the scattering can be neglected,
the comoving free-streaming scale at the matter-radiation equality
$R_f$ is given as
\begin{eqnarray}
R_f &=& \int^{t_{EQ}}_{t_I} \frac{v(t')}{a(t')} dt'
\nonumber\\
    &\simeq& 2 v_0 t_{EQ} (1+z_{EQ})^2 
\log\left(\sqrt{1+\frac{1}{v_0^2 (1+z_{EQ})^2}} 
          +\frac{1}{v_0 (1+z_{EQ})} \right)
\label{rf}
\end{eqnarray}
where $z_{EQ}$ and $t_{EQ}$ are the red shift and cosmic time for the
matter-radiation equality \cite{Borgani:1996ag}. The $v_0$ is the
current velocity of the LSP,
\begin{eqnarray}
v_0= \frac{T_0}{T_I} \frac{E_I}{m_{\tilde{\chi}^0_1}}
\end{eqnarray}
where $T_0$ and $T_I$ are temperatures for the current cosmic
microwave background radiation and the production time of the LSP,
$E_I$ and $m_{\tilde{\chi}^0_1}$ are the energy at the production time
and the mass for the LSP. In order to explain the small-scale
structure of $O(0.1)$ Mpc well, $v_0$ is preferred to be $10^{-(7-8)}$
from Eq.~(\ref{rf}), and this means \cite{Lin:2000qq}
\begin{eqnarray}
\frac{E_I}{T_I} = 
2.1 \times 10^7 \times
\left(\frac{m_{\tilde{\chi}^0_1}}{50{\rm GeV}}\right)
\left(\frac{v_0}{10^{-7}}\right).
\label{eovert}
\end{eqnarray}

Other energetic particles are likely associated with the LSP
production. Therefore we mainly consider a case where $T_I$ at the
production time is larger than about a few MeV so that the standard
neucleosynthesis works. We will come back to this point
later.\footnote{
One might worry that the LSP is relativistic at the neucleosynthesis era
and it may change the expansion rate significantly.
Assuming that the LSP is the dark matter of the
universe and Eq.~(\ref{eovert}), the energy density of the LSP at the
neucleosynthesis era is $\sim 0.2\% (v_0/10^{-7})$ of that of three
neutrinos, and it does not give any significant effect on the
neucleosynthesis.  
}

So far we assumed that the LSP does not lose its relativistic energy
significantly in the scattering processes in the thermal bath.  If the
LSP is gravitino or axino, it does not lose its energy by the
scattering because it couples with particles of the SUSY SM very
weakly \cite{Borgani:1996ag}. However, the neutralino LSP is weak
interacting, and it may lose most of its energy by the scattering
processes in thermal bath. In this letter we calculate the energy
reduction in the successive scattering of the relativistic LSP in the
thermal bath. We find the energy reduction is suppressed and the
neutralino could be the WDM if some conditions are satisfied. For
$v_0$ is $10^{-(7-8)}$, $T_I$ can not exceed over $\sim 5$MeV for the
bino-like LSP, $\sim 2$MeV for the Higgsino-like LSP. The wino-like
LSP cannot be the WDM.

First, we review  nature of the neutralino LSP. The neutralinos are
composed of bino, wino, and two Higgsinos. The mass matrix is
\begin{eqnarray}
   M_N&=&
   \left(
   \begin{array}{cccc}
     M_{\tilde{B}}    & 0 & -m_Z s_W c_\beta & m_Z s_W s_\beta \\
     0 & M_{\tilde{W}} & m_Z c_W c_\beta & -m_Z c_W s_\beta \\
     -m_Z s_W c_\beta & m_Z c_W c_\beta & 0 & -\mu
     \\ 
     m_Z s_W s_\beta & -m_Z c_W s_\beta & -\mu & 0 
   \end{array}            \right).
\end{eqnarray}
Here $M_{\tilde{B}}$, $M_{\tilde{W}}$, and $\mu$ are the bino, wino,
and supersymmetric Higgsino masses, respectively.
$c_\beta(\equiv\cos\beta)$ and $s_\beta(\equiv\sin\beta)$ are for a
mixing angle of the vacuum expectation values of the Higgs bosons, and
$c_W(\equiv\cos\theta_W)$ and $s_W(\equiv\sin\theta_W)$ for the
Weinberg angle. If $M_{\tilde{B}}\ll \mu, M_{\tilde{W}}$, the LSP is
bino-like. On the other hand, if $M_{\tilde{W}}$ or $\mu$ is smaller
than the others, the LSP is wino- or Higgsino-like. In the wino- and
Higgsino-like cases, the LSP and the lighter chargino are degenerated in
masses.  The next lightest SUSY particle would be important for our
discussion since the inelastic scattering of the LSP contributes to
the energy reduction.

In the minimal supergravity model, the bino-component is dominant in
the LSP.  This is because $M_1\sim 0.5 M_2$ and $\mu$ tends to be
larger than the gaugino masses due to the radiative breaking
condition. However, if the universal gaugino mass condition at the
gravitational scale is broken, LSP can be wino-like. Especially, in
the anomaly mediation SUSY breaking model, the wino-component
dominates over the others since the gaugino masses are proportional to
the one-loop beta function of the gauge coupling constants in the SUSY
SM \cite{anomaly}. For a very large universal scalar mass compared to
the gaugino masses in the minimal supergravity model
\cite{Feng:2000zu} or breakdown of universality of the scalar masses
may lead to the Higgsino-like LSP. In this letter we do not assume any
specific SUSY breaking models and discuss each the neutralino LSPs.

The energy loss of the relativistic LSP depends on the temperature at
the LSP production time, $T_I$.  If the LSP is produced below $T_C$,
\begin{eqnarray}
T_C = 6.3 {\rm MeV}
\left(\frac{m_{\tilde{\chi}_1^0}}{50{\rm GeV}}\right)^{1/2}
\left(\frac{v_0}{10^{-7}} \right)^{-1/2},
\end{eqnarray}
it is  typically non-relativistic in the CM frame of the scattering
processes with particles in the thermal bath. In this case the energy
reduction par one scattering $r(\equiv \Delta E/E)$ is
\begin{eqnarray}
    r &=& 4 \frac{q E}{m^2_{\tilde{\chi}^0_1}}\sin^2{(\theta/2)}
    \sin^2{(\eta/2)}.
\label{r_nr}
\end{eqnarray}
Here, $q$ is the energy of a particle in the thermal bath, which is
$\sim 3T$, and $\theta$ is the relative angle between the LSP and the
particle in the thermal bath, and $\eta$ the scattering angle of the
LSP in the CM frame. Here we take a leading term of $O(q/E)$. Then,
the energy reduction is suppressed by $O(T E/m^2_{\tilde{\chi}^0_1})$
when $T\lsim T_C$. On the other hand, if the LSPs are produced above $T_C$,
they are relativistic in the CM frame of the scattering processes at
the production time, and the energy reduction is unsuppressed as
\begin{eqnarray}
r &=& \sin^2{(\eta/2)}. 
\end{eqnarray}
Provided that the event rate is faster than the Hubble expansion, the
LSP loses the energy quickly so that LSP scattering becomes
non-relativistic in the CM frame.

First, we consider the case where $T_I \lsim T_C$ and calculate the energy
reduction of the LSP due to the two-body elastic scattering in the
thermal bath. The evolution of the LSP energy is given as
\begin{eqnarray}
\label{eq:dedt}
\frac{d E}{d t} = 
-H E 
-\sum_i g_i \int \frac{d^3 q}{(2 \pi)^3}~ {\rm e}^{-\frac{q}{T}}~ 
(r E)~ v_{rel} \frac{d \sigma_i}{d r} dr.   
\end{eqnarray}
Here $H$ is the Hubble parameter. The index $i$ is for spices of
particle in the thermal bath with the degrees of freedom $g_i$,
$v_{rel}$ and $\sigma_i$ are the relative velocity and the cross
section of the elastic scattering between the LSP and the particle in
the thermal bath.  Since the LSP is neutral and $T_I$ is smaller than
$T_C$ and larger than 1MeV, $i=e^-$, $\nu_e$, $\nu_\mu$, $\nu_\tau$,
and the anti-particles. The contributing diagrams to the energy
reduction come from the $Z$ boson and slepton exchanges.  We will take
a massless limit for the particles in the thermal bath for simplicity.
The explicit calculation gives
\begin{eqnarray}
\sum_i g_i \int \frac{d^3 q}{(2 \pi)^3}~ {\rm e}^{-\frac{q}{T}} 
(r E)~ v_{rel} \frac{d \sigma_i}{d r} dr   
&=&
\frac{16}{\pi^3} \left(|A_L^{(i)}|^2 +|A_R^{(i)}|^2 \right)
\frac{E^4 T^6}{m^4_{\tilde{\chi}^0_1}}.
\label{explicit}
\end{eqnarray}
Here,
\begin{eqnarray}
A_L^{(e)} &=& 
\frac{g_2^2}{m_Z^2 c^2_W}C_{11} L_e 
-
\frac{g_2^2}{2 m_{\tilde{e}_L}^2} ([O_N]_{12}+ [O_N]_{11} t_W)^2,
\nonumber\\
A_R^{(e)} &=& 
\frac{g_2^2}{m_Z^2 c^2_W}C_{11} R_e 
+
\frac{2 g_2^2}{m_{\tilde{e}_L}^2} ([O_N]_{11} t_W)^2,
\nonumber\\
A_L^{(\nu)} &=& 
\frac{g_2^2}{m_Z^2 c^2_W}C_{11} L_\nu 
-
\frac{g_2^2}{2 m_{\tilde{\nu}_L}^2} ([O_N]_{12}- [O_N]_{11} t_W)^2,
\nonumber\\
A_R^{(\nu)} &=& 
0
\nonumber
\end{eqnarray}
where $C_{11} = ([O_N]_{13}^2-[O_N]_{14}^2)$ with $[O_N]$ the
diagonalization matrix of $M_N$, $L_i= T_3+Q s_W^2$, $R_i=Q s_W^2$,
and $t_W\equiv\tan\theta_W$. The momentum transfers on the propagators
of the exchanged particles are negligible compared with the masses,
thus we replace the propagators of $Z$ boson and sleptons to their
mass squares $m^2_Z$ and $m^2_{\tilde{l}}$, respectively.  By solving
Eq.~(\ref{eq:dedt}), the LSP energy at the radiation-matter equality
is given as
\begin{eqnarray}
{E_{EQ}} &=& E_I\left(\frac{{T_{EQ}}}{T_I}\right)
\left(1- \left(\frac{\Delta E}{E}\right)_{\rm eff} \right),
\label{formula}
\end{eqnarray}
where
\begin{eqnarray}\label{tyon}
\left(\frac{\Delta E}{E}\right)_{\rm eff} 
&=&
\frac{24 \sqrt{5}}{7 \pi^{\frac92}}
{g_*}^{-\frac12} 
\sum_i \left(|A_L^{(i)}|^2 +|A_R^{(i)}|^2 \right) 
\frac{M_{pl} E_I^3 T_I^4}{m^4_{\tilde{\chi}^0_1}}. 
\end{eqnarray}
Here, $g_*$ is total number of the effective degrees of freedom for at
the temperature $T_I$. We assume that the universe is radiation
dominant and use $H= (4 \pi^3/45)^{1/2} {g_*}^{\frac12} T^2/M_{pl}$
for the Hubble parameter.  The first bracket in the right-handed side
in Eq.~(\ref{formula}) comes from the red-shift due to the expansion
of the universe, and the second one is the effect from the scattering
of the LSP in the thermal bath. Here we expand $E_{EQ}$ by
$E_IT_I/m^2_{\tilde{\chi}^0_1}$ and keep the leading term in
Eq.~(\ref{tyon}), assuming the energy reduction from the scattering is
small. When $(\Delta E/E)_{\rm eff}$ is larger than one, $T_I$
is replaced to the temperature at which the elastic scattering becomes
ineffective to the LSP energy reduction, and $E_I$ is  given by
the LSP energy at the $T_I$. This means that our result is conservative.

In Eq.(\ref{tyon}), $(\Delta E/E)_{\rm eff}$ is suppressed by
$T^4_{I}$. This comes from the suppression in the amplitude and the
phase space, in addition to the energy reduction in the
non-relativistic limit of the LSP (Eq.~(\ref{r_nr})). The momentum
transfers in the scattering processes ($\sim E T$) are smaller than
the exchanged particle masses in the amplitude, and the phase space of
the elastic scattering is also suppressed by
$ET/m^2_{\tilde{\chi}^0_1}$. Thus, the event rate par a Hubble time is
smaller in lower temperature by $\propto T^3$ as
\begin{eqnarray}
\frac{\Gamma}{H} &\equiv&
\frac{1}{H} \sum_i g_i \int \frac{d^3 q}{(2 \pi)^3}~ {\rm e}^{-\frac{q}{T}} 
~ v_{rel} \sigma_i  
\nonumber\\
&=&
\frac{45\sqrt{5}}{16\pi^{\frac92}}
{g_*}^{-\frac12} 
\sum_i \left(|A_L^{(i)}|^2 +|A_R^{(i)}|^2 \right) 
\frac{M_{pl} E^2 T^3}{m^2_{\tilde{\chi}^0_1}}.
\label{gammah}
\end{eqnarray}
Note that the energy reduction $({\Delta E}/{E})_{\rm eff}$ is dominated
by the contribution at $T = T_I$ and is not sensitive to $T_{EQ}$.

If the LSP is bino-like, the $Z$ boson exchange contribution is
suppressed by $m_Z^2/\mu^2$ in the amplitude. Then, the slepton
exchange contribution dominates if $\mu$ is larger than the slepton
masses. Taking the pure bino limit,
\begin{eqnarray}
\left(\frac{\Delta E}{E}\right)_{\rm eff} 
&=& 
3.9\times 10^{-2} 
\left(\frac{m_{\tilde{\chi}_1^0}}{50{\rm GeV}}\right)^{-1}
\left(\frac{m_{\tilde{l}}}{1{\rm TeV}}\right)^{-4}
\left(\frac{v_0}{10^{-7}}\right)^{3}
\left(\frac{T_I}{1{\rm MeV}}\right)^{7}.
\label{binosl}
\end{eqnarray}
Here $m_{\tilde{\chi}_1^0}\simeq M_{\tilde{B}}$, and we take
$m_{\tilde{e}_R}=m_{\tilde{e}_L}=m_{\tilde{\nu}_L}(\equiv
m_{\tilde{l}})$.  In order to suppress the energy reduction below the
10\% so that the LSP can behave as the WDM, $T_I$ should be smaller
than 1.1(3.1)MeV for $v_0=10^{-7}(10^{-8})$,
${m_{\tilde{\chi}_1^0}}=50$GeV and $m_{\tilde{l}}<1$ TeV. This value
corresponds to $E_I=$24(6.5)TeV from Eq.~(\ref{eovert}). If the LSP is
heavier, the energy reduction is suppressed more, and a slightly
larger $T_I$ is possible.  For ${m_{\tilde{\chi}_1^0}}=200$GeV, $T_I$
should be smaller than 1.4(3.7)GeV for $v_0=10^{-7}(10^{-8})$. This
means that $E_I<$ 118(32)TeV.

Note that calculation of the energy reduction rate is valid only when
$\Gamma/H > 1$ .  In the bino dominant limit, the event rate of the
elastic scattering process by the slepton exchange par a Hubble time
is
\begin{eqnarray}
\frac{\Gamma}{H} &=&
3.8
\left(\frac{m_{\tilde{l}}}{1{\rm TeV}}\right)^{-4}
\left(\frac{v_0}{10^{-7}}\right)^{2}
\left(\frac{T_I}{1{\rm MeV}}\right)^{5}.
\end{eqnarray}
The event rate is not still sufficiently suppressed compared
with the Hubble expansion.

In Eq.~(\ref{binosl}) we took the slepton masses 1TeV. However, some
SUSY breaking models predict much heavier sleptons, which is not
necessarily in conflict with the naturalness argument \cite{es}.  When
sfermions are heavy, the thermal component of the bino-like LSP cannot
annihilate sufficiently in the thermal processes so that the energy
density might be too large beyond the critical density. However, if
the huge entropy  is supplied in the non-thermal process as
mentioned before, it can be diluted and be harmless. 

When the slepton exchange is sufficiently suppressed, the $Z$ boson
exchange becomes dominant in the energy reduction of the LSP. The
energy reduction by the $Z$ boson exchange is given as
\begin{eqnarray}
\left(\frac{\Delta E}{E}\right)_{\rm eff} 
&=& 
6.9\times 10^{-3}
\left(\frac{m_{\tilde{\chi}_1^0}}{50{\rm GeV}}\right)^{-1}
\left(\frac{\mu}{1 {\rm TeV}}\right)^{-4}
\left(\frac{v_0}{10^{-7}}\right)^{3}
\left(\frac{T_I}{1{\rm MeV}}\right)^{7} \cos^2 2\beta.
\label{Zcont}
\end{eqnarray}
Here we used the approximated solution $C_{11}=-m_Z^2 s_W^2 \cos 2
\beta/\mu^2 $ for $m_Z$, $ M_{\tilde{B}}\ll\mu$.  Recent LEP II
searches of the light Higgs boson prefer  $\vert \cos 2\beta\vert
>0.53$~\cite{higgs}. Since the Higgsino mass $\mu$ is related with the
Higgs boson mass, we cannot take too large a value for $\mu$ compared
to the weak scale from the naturalness argument. From
Eq.~(\ref{Zcont}), $T_I$ should be smaller than 5MeV assuming $\mu$ is
smaller than 1TeV, ${m_{\tilde{\chi}_1^0}}={200{\rm GeV}}$, and
${v_0}={10^{-8}}$.

Next, let us consider the Higgsino-like LSP. In this case the slepton
exchange contribution is suppressed by the small gaugino components
and the Yukawa coupling constants, and the $Z$ boson exchange
contribution dominates in the elastic scattering processes. Then,
\begin{eqnarray}
\left(\frac{\Delta E}{E}\right)_{\rm eff} 
&=& 
2.4
\left(\frac{m_{\tilde{\chi}_1^0}}{100{\rm GeV}}\right)^{-3}
\left(\frac{m_{\tilde{B}}}{500{\rm GeV}}\right)^{-2}
\left(\frac{v_0}{10^{-7}}\right)^{3}
\left(\frac{T_I}{1{\rm MeV}}\right)^{7}
\cos^22\beta.
\label{higgsino}
\end{eqnarray}
Here, $m_{\tilde{\chi}_1^0}\simeq \mu$ and it should be larger than
about 100GeV from negative search for the Higgsino-like chargino.
When the gaugino masses are heavier than $\mu$ and $m_Z$, $C_{11}$ is
given as
\begin{eqnarray}
C_{11} &=& \mp \frac{m_Z^2}{2 \mu}
\left(\frac{s_W^2}{m_{\tilde{B}}}+\frac{c_W^2}{m_{\tilde{W}}}\right)
\cos2\beta,
\end{eqnarray}
for $\mu$ positive (negative).   It further reduces to 
\begin{eqnarray}
C_{11} &=& \mp \frac43 \frac{m_Z^2 s_W^2}{\mu m_{\tilde{B}}}\cos2\beta
\end{eqnarray}
by using the GUT relation $M_{\tilde{B}}/M_{\tilde{W}}=5/3 t_W^2\simeq
1/2$.  In Eq.~(\ref{higgsino}) we used this formula for simplicity. In
order to suppress the energy reduction by the elastic scattering,
$T_I$ should be smaller than 0.85(2.3)MeV for $v_0=10^{-7}(10^{-8})$
and a relatively heavy LSP mass ${m_{\tilde{\chi}_1^0}}=200$GeV as far
as the gaugino masses are smaller than 1TeV.

We saw that  the elastic scattering of the Higgsino-like LSP is suppressed
for heavier gaugino masses. However, we have to check if the inelastic
scattering of the LSP by the $W$ boson exchange does not contribute to
the energy reduction. As we noted before, the chargino is degenerate
with LSP in masses. The Boltzmann suppression, $\exp(-
(m_{\tilde{\chi}_1^0} \Delta m_{\tilde{\chi}})/2ET)$, may not be
too small.  Furthermore the coupling with $W$ boson is not suppressed
at all. Therefore the processes, such as $\tilde{\chi}_1^0 e^-
\rightarrow \tilde{\chi}_1^- \nu$, may be important over the Boltzmann
suppression factor.

 The mass difference between the chargino and the LSP is
\begin{eqnarray}
\Delta m_{\tilde{\chi}} &\equiv& (m_{\tilde{\chi}_1^+}-m_{\tilde{\chi}_1^0})
\nonumber\\
&=&
\left(
\frac{1\pm \sin2\beta}{2} \frac{m_Z^2 s_W^2}{M_{\tilde{B}}}
+
\frac{1\mp \sin2\beta}{2} \frac{m_Z^2 c_W^2}{M_{\tilde{W}}}
\right)
\\
&=&
\frac{m_Z^2 s_W^2}{M_{\tilde{B}}}
\left(\frac43 \mp\frac13 \sin 2\beta\right),
\end{eqnarray}
and this is about 5GeV for $M_{\tilde{B}}=$500GeV. The energy
reduction by one scattering of $\tilde{\chi}_1^0 e^- \rightarrow
\tilde{\chi}_1^- \nu$ is
\begin{eqnarray}
r &=& 4 \frac{q E}{m^2_{\tilde{\chi}^0_1}}\sin^2{(\theta/2)} \sin^2{(\eta/2)}
 -2 \frac{\Delta m_{\tilde{\chi}}}{m_{\tilde{\chi}^0_1}}.
\end{eqnarray}
From the kinematics, $r$ is positive definite. Each chargino decay
also reduces the energy of the order of $\Delta
m_{\tilde{\chi}}/m_{\tilde{\chi}_1^0}$.  The event rate of the inverse
inelastic scattering processes of chargino, such as $\tilde{\chi}_1^-
\nu_e \rightarrow \tilde{\chi}_1^0 e^-$, is suppressed by $120\pi
(T/\Delta m_{\tilde{\chi}})^3$ compared with the decay rate, thus
contribution to the energy reduction is negligible.

The event rate of the inelastic scattering of the Higgsino-like LSP
par a Hubble time is
\begin{eqnarray}
\frac{\Gamma}{H} 
&=&
\frac{3 \sqrt{5}}{4 \pi^{\frac32}} {g_*}^{-\frac12} 
g_2^4 \frac{M_{pl} E T^2}{m^4_W}
{\rm e}^{- \frac{m_{\tilde{\chi}_1^0} \Delta m_{\tilde{\chi}}}{2ET}}
\left( 
\frac{\Delta m_{\tilde{\chi}}}{m_{\tilde{\chi}_1^0}} +
6 \frac{ET}{m_{\tilde{\chi}_1^0}^2} 
\right) N_F
\end{eqnarray}
where $N_F$ is the number of the inelastic processes.  When the mass
difference is larger than
\begin{eqnarray}
\frac{2 T_I E_I}{m_{\tilde{\chi}_1^0}} &=&
850 {\rm MeV} 
\left(\frac{v_0}{10^{-7}}\right)
\left(\frac{T_I}{1{\rm MeV}}\right)^{2},
\end{eqnarray}
the inelastic processes are suppressed by the Boltzmann factor. Then,
${\Gamma}/{H}$ is sensitive to $T$, $v_0$, and $\Delta
m_{\tilde{\chi}}$. If $v_0=10^{-7}$ and $\Delta m_{\tilde{\chi}} =
5$GeV, ${\Gamma}/{H}$ is of the order of $10^5$ even for $T=1$MeV, and
the energy reduction by the inelastic scattering cannot be suppressed.
On the other hand, if $v_0=10^{-8}$ and $\Delta m_{\tilde{\chi}} =
5$GeV, it is $10^{-20}$ (33) for $T=1(2)$MeV, and the energy reduction
may be suppressed. The Higgsino-like WDM is marginally viable.

Note that the energy of the chargino produced by the inelastic
scattering is also reduced by the electromagnetic interaction. The
life time of the Higgsino-like chargino is
\begin{eqnarray}
\tau_{\tilde{\chi}_1^-}^{-1} 
&=&
N_D \frac{g_2^4}{960 \pi^3} \frac{{\Delta m_{\tilde{\chi}}}^5}{m_W^4} 
\end{eqnarray}
with $N_D$ the number of the decay modes.  The energy reduction by the
electromagnetic interaction is given as
\begin{eqnarray}
\frac{dE}{dt} = -\frac{\pi^3 \alpha^2}{3} \Lambda T^2
\end{eqnarray}
where $\Lambda$ is of the order of 1 \cite{reno}. Then, the energy
reduction rate of the Higgsino-like chargino in one life is
\begin{eqnarray}
\left(\frac{\Delta E}{E}\right)_{\rm 1-life} 
\sim 1.3 \times 10^{-3}~
N_D  \Lambda 
\left(\frac{T}{1{\rm MeV}}\right)^2 
\left(\frac{m_{\tilde{\chi}_1^0}}{100{\rm GeV}}\right)^{-1} 
\left(\frac{\Delta m_{\tilde{\chi}}}{5{\rm GeV}}\right)^{-5}.
\end{eqnarray}
This effect may be harmless if $\Delta m_{\tilde{\chi}} =
5$GeV.

When the LSP is wino-like, the elastic scattering can be suppressed if
the slepton and the Higgsino masses are heavy, similar to the
bino-like LSP. However, when the $Z$ boson contribution is suppressed
by raising the Higgsino mass, the chargino and the LSP become more
degenerate in masses than in the Higgsino-like case as
\begin{eqnarray}
\Delta m_{\tilde{\chi}}
&=&
\frac{m_Z^4}{M_{\tilde{B}} \mu^2} s_W^2 c_W^2 \sin^22\beta
\end{eqnarray}
for $M_{\tilde{W}}$, $m_Z\ll M_{\tilde{B}}$, $\mu$. If $\mu$ is 1TeV
and $M_{\tilde{B}}$ is 100GeV, $\Delta m_{\tilde{\chi}}$ is about
100MeV. The $\Gamma/H$ for the inelastic scattering by the $W$ boson
exchange is of the order of $10^5$ for $\Delta
m_{\tilde{\chi}}=100$MeV even if $T=1$MeV and $v_0=10^{-8}$. Since
either the $Z$ or $W$ boson exchange contributions cannot be
suppressed, the wino-like LSP cannot be the WDM.

Finally, we discuss the case for $T_I \gsim T_C$. In this case, the
momentum on the exchanged particle is not negligible, and the event
rate becomes larger than in the case of the lower
temperature. Therefore the energy reduction becomes maximum at $T
\simeq T_C$ and the LSP loses the relativistic energy till the
temperature goes down to $T_C$.

As an example, we present the energy reduction of the bino-like LSP by
the elastic scattering since the constraint on the $T_I$ is the
weakest among the neutralino LSPs. Assuming the slepton exchange is
suppressed by the heavy masses, the $Z$ boson contribution to the
energy reduction when the typical momentum transfer is much larger
than $m_Z^2$ ($ET\gg m^2_Z$) is expressed by
\begin{eqnarray}
    \sum_i g_i \int \frac{d^3 q}{(2 \pi)^3}~ {\rm e}^{-\frac{q}{T}} (r
    E)~ v_{rel} \frac{d \sigma_i}{d r} dr &=& \sum_{i}\zeta\frac{g_2^4
    t^4_W}{64\pi^3} (L_i^2+R_i^2) \frac{m_Z^4}{\mu^4} \cos^22\beta~
    T^2,
\end{eqnarray}
where $\zeta= (2 \log\frac{4ET}{m_Z^2}-5 -2 \gamma).$
The energy reduction rate is given as 
\begin{eqnarray}
\left(\frac{\Delta E}{E}\right)_{\rm eff} 
&=&
1.0 \times 10^3 \zeta
\left(\frac{m_{\tilde{\chi}_1^0}}{50{\rm GeV}}\right)^{-1}
\left(\frac{\mu}{1 {\rm TeV}}\right)^{-4}
\left(\frac{v_0}{10^{-7}}\right)
\left(\frac{T_C}{100{\rm MeV}}\right)^{-1} \cos^2 2\beta,
\end{eqnarray}
and it is difficult for the LSP to keep the relativistic energy.

In this letter we calculate the energy reduction of the LSP which is
produced by the non-thermal process and study whether the LSP can be
the warm dark matter or not. If the temperature of the production time
$T_I$ is smaller than 5MeV, the bino-like LSP can be the WDM and may
contribute to the small-scale structure of $O(0.1)$ Mpc. The
Higgsino-like LSP might also work as the WDM if $T_I<$ 2MeV. The
wino-like LSP cannot be the WDM.

We now discuss the some of the aspects on the mechanism to produce
relativistic neutralino. Here we discuss the LSP produced from heavy
moduli decay. Such a moduli might dominate the energy density of
universe before its decay. Therefore the moduli decay associates with
the large entropy production and reheating. For this case, the LSP
energy density over the entropy density at present might be written as
\begin{equation}
    m_{\tilde{\chi}_1^0}Y_{\tilde{\chi}_1^0} \simeq 0.75 \times10^{-6}
    {\rm GeV}~ \bar{N}_{\tilde{\chi}_1^0}
    \left(\frac{m_{\tilde{\chi}_1^0}}{100 {\rm GeV}} \right)
    \left(\frac{T_I}{1{\rm MeV}}\right) \left( \frac{m_{\phi}}{100
      {\rm TeV}}\right)^{-1}.
\end{equation}
Here we identify $T_I\equiv T_R$, $\bar{N}_{\tilde{\chi}_1^0}$ is
average number of the LSP from a moduli decay. On the other hand,
$m_{\tilde{\chi}_1^0}Y_{\tilde{\chi}_1^0}\sim 10^{-9}$ GeV is
preferred as the dark matter density. This leads
$\bar{N}_{\tilde{\chi}_1^0}$ must be a order of $10^{-3}$. Such a
small branching ratio is expected for the case where the moduli decay
into gravitino is suppressed \cite{mr}. The small branching ratio also
means that many energetic particles are produced associated with the
moduli decay, thus the effect on the neucleosynthesis must be
considered. For $T_R\gsim 2.5$--$4$ MeV such an effect would be small
enough \cite{Kawasaki:2000en}. On the other hand, if the energy
density of the heavy moduli does not exceed over that of one neutrino
spices at the time of neucleosynthesis and it decays before $10^{4}$ sec,
the associated high energetic particles are thermalized before they
hits to nuclei, provided that the hadrons are not produced in the
decay.  In such case, the decay can be harmless to the
neucleosynthesis.

For the neutralino LSP to stay warm and produced above 1 MeV, the LSP
must be either nearly pure bino or Higgsino, in order to suppress the
scattering in the thermal bath. This means counting rate at the
conventional dark matter detectors would be very small. Discovery of
the dark matter signal in any forthcoming experiments \cite{CDMS} will
suggest the LSP is not the warm dark matter. For the bino-like LSP the
slepton masses also need to be very heavy. If deviation of the muon
anomalous magnetic moment from the standard model prediction is
observed, the warm bino-like LSP is disfavoured \cite{gminus}.

\underline{Note Added:} After completion of this work, there appears a
paper where the energy reduction of the WIMP by scattering in the
thermal bath is also discussed\cite{mac}. They assume 
the WIMP is produced by the
non-thermal process, but the WIMP
is not the LSP. Also, they impose $\Gamma/H<1$ for the scattering
processes, and do not calculate the energy reduction rate of
the WIMP by scattering.

\section*{Acknowledgement}
We would like to thank R.~Brandenberger, L.~Roszkowski, and Y.~Suto for
useful discussions. This work was supported in part by the
Grant-in-Aid for Scientific Research from the Ministry of Education,
Science, Sports and Culture of Japan, on Priority Area 707
``Supersymmetry and Unified Theory of Elementary Particles" (J.H.)
and Grant-in-Aid for Scientific Research from the Ministry of
Education (12047217, M.M.N) 

\newpage
%
%
%
\newcommand{\Journal}[4]{{\sl #1} {\bf #2} {(#3)} {#4}}
\newcommand{\PL}{\sl Phys. Lett.}
\newcommand{\PR}{\sl Phys. Rev.}
\newcommand{\PRL}{\sl Phys. Rev. Lett.}
\newcommand{\NP}{\sl Nucl. Phys.}
\newcommand{\ZP}{\sl Z. Phys.}
\newcommand{\PTP}{\sl Prog. Theor. Phys.}
\newcommand{\NC}{\sl Nuovo Cimento}
\newcommand{\MPL}{\sl Mod. Phys. Lett.}
\newcommand{\PRep}{\sl Phys. Rep.}

\end{document}